# Efficiency of research performance and the glass researcher

Lutz Bornmann* & Robin Haunschild**

*Division for Science and Innovation Studies

Administrative Headquarters of the Max Planck Society

Hofgartenstr. 8,

80539 Munich, Germany.

Email: bornmann@gv.mpg.de

**Max Planck Institute for Solid State Research

Heisenbergstr. 1,

70569 Stuttgart, Germany.

**Abstract**

Abramo and D'Angelo (in press) doubt the validity of established size-independent indicators measuring citation impact and plead in favor of measuring scientific efficiency (by using the Fractional Scientific Strength indicator). This note is intended to comment on some questionable and a few favorable approaches in the paper by Abramo and D'Angelo (in press).

Abramo and D'Angelo (in press) doubt the validity of established size-independent indicators for measuring citation impact. In order to demonstrate the supposed lack of validity some examples are presented in their paper which are based on data from a comprehensive national database. The database contains standardized data on the Italian science system which includes not only output but also input indicators. Abramo and D'Angelo (in press) plead in favor of measuring scientific efficiency rather than performance, and refer to their proposal of the Fractional Scientific Strength (FSS) indicator (Abramo & D'Angelo, 2014). This is a composite indicator which considers – when used on the university level – the total salary of the research staff and the total number of publications which are weighted with citation impact. The argumentations and recommendations of Abramo and D'Angelo (in press) may sound reasonable from an economic standpoint, but are questionable in other respects. In the following, we take up some problematic and two interesting points.

**Some problematic points**

(1) In the beginning of the paper, Abramo and D'Angelo (in press) question the validity of size-independent citation impact indicators for measuring research performance. They argue that research performance can be validly measured only if input indicators are considered (e.g. citation impact per euro spent or FSS). There are several problems with this approach. (i) Scientometricians do not equate citation impact to research performance. The measurement of research performance includes several aspects other than impact alone. For example, the SCImago Institutions Ranking (http://www.scimagoir.com/) considers manifold sets of indicators which are categorized as research indicators (e.g. output and scientific talent pool), innovation indicators (e.g. technological impact), or web visibility indicators (e.g. website size). (ii) Abramo and D'Angelo (in press) confuse impact with efficiency and performance with productivity. They claim that size-independent indicators violate an axiom of production theory, but size-independent impact indicators were never intended to measure productivity. (iii) The plea of Abramo and D'Angelo (in press) for efficiency scores and against citation impact indicators is based on "basic economic reasoning". The authors fail to substantiate their claim that one metric measures research performance validly and the others do not. The psychometric literature offers several ways to investigate validity which have already been applied in bibliometrics: the investigation of predictive validity studies the degree to which an indicator can predict other indicators of research performance which are measured in the future (Hirsch, 2007). For the investigation of the convergent validity, it is studied whether an indicator is correlated with other indicators which are specified to measure the same aspect (Bornmann, Mutz, & Daniel, 2008). However, the investigation of validity requires the comparison of different indicators measuring the same construct of performance. Since citation indicators measure impact only and composite indicators (the FSS) consider several aspects of research performance; they do not claim to measure the same construct and cannot be fairly compared. In order to test the predictive or convergent validity of the FSS, Abramo and D'Angelo (in press) need at least one other indicator which measures research performance validly. Then, they can really judge the validity of the FSS.

(2) Abramo and D'Angelo (in press) argue for the use of a composite indicator to measure research performance. The indicator considers some input and output variables which are combined into a single number. Instead, we argue that research performance should not be measured by composite indicators, but by reporting the results of measuring different aspects of research performance separately. Research performance is a multi-dimensional phenomenon and different dimensions should be reported in an evaluation study. For example, Bornmann and Marx (2014) proposed using 16 indicators (each of which is more or less important in a specific evaluation) to measure the productivity and citation impact of single researchers. As input measures one can use salaries, but also requested external funds or number of doctoral and post-doctoral researchers (in a professorship or institution). Moed and Halevi (2015) introduce the Multidimensional Research Assessment Matrix. The matrix can be used to decide which indicators are applied in a specific evaluation context. The basic assumption is that "the choice of metrics to be applied in a research assessment process depends on the unit of assessment, the research dimension to be assessed, and the purposes and policy context of the assessment" (Moed & Halevi, 2015). In order to study research performance in different contexts validly, we recommend following flexible schemes which can be adapted to a specific evaluation context rather than fixed formulas which try to combine a small selection of variables in a composite indicator. This is aggravated by the fact that with the use of composite indicators the included variables remain black boxes.

(3) Abramo and D'Angelo (in press) use fictitious examples where John Doe with 70 publications performs better than Jane Doe with 1 publication according to FSS. The situation is vice versa according to MNCS. Of course, the size-independent indicators offer one perspective which has to be supplemented with another perspective, the size-dependent one. This does not mean that one of the two perspectives is valid and the other is not. Both are needed for a comprehensive research evaluation (Bornmann & Haunschild, in press; Haunschild & Bornmann, 2015).

(4) The fourth point is related to the second. The selection of indicators which are used in an evaluation should also consider the availability of data for international comparisons. Giovanni Abramo and his co-authors use data on researchers and institutions in Italy extracted from a database on Italian university personnel, maintained by the Ministry of Education, Universities and Research. As described in various papers (e.g. Abramo, Cicero, & D'Angelo, 2013; Abramo & D'Angelo, 2011), the database offers normalized citations (NCs) for publications. The NCs are calculated by dividing the citation counts of a focal paper by the average number of citations of all *Italian* publications of the same document type, publication year, and Web of Science (WoS) subject category. This is a specific calculation of NCs, because only Italian publications are used in the denominator, and does not follow the standards in bibliometrics where all publications are used (of the same document type, publication year, and WoS subject category). Since international comparison is the objective in most research evaluations, evaluation techniques which are based on country-specific data cannot be the grand solution. The national science systems are so different that they offer only a restricted set of indicators which can be used in an international context. The national NCs, salaries of researchers, and research funds available are examples of indicators which can hardly be used to evaluate research performance internationally.

(5) Abramo and D'Angelo (in press) plead for complex databases – similar to the Italian database – which include various input and output indicators on the level of single researchers. Since we are not only scientometricians (who appreciate the availability of data), but also researchers ourselves, we should not support such transparent systems which form the "glass researcher". The systems are an invitation for politicians and decision makers to combine – more or less competently – various input and output indicators for an efficiency measurement. Furthermore, we should ask ourselves in a general sense whether we should force the use of efficiency metrics in research evaluation processes. Creativity and faulty incrementalism are basic elements of the research process, which are diametrically opposed to efficiency. According to Ziman (2000) "the post-academic drive to 'rationalize' the research process may damp down its creativity. Bureaucratic 'modernism' presumes that research can be directed by policy. But policy prejudice against 'thinking the unthinkable' aborts the emergence of the unimaginable" (p. 330). Scientometricians should try to explore methods to measure the efficiency of research, but they should be very careful in using the methods in evaluation studies which are used for decision making. While such methods might be used in evaluation studies about applied science, they should not be used to evaluate fundamental sciences.

**Two interesting points**

Despite our critical view of the paper by Abramo and D'Angelo (in press) we would like to mention two interesting points raised by the authors. (i) The formula of the FSS sums NC instead of averaging it (as normally done). Thus, this part of the formula combines impact and output by weighting each publication with its citation impact. The approach follows earlier approaches like the I3 indicator (Leydesdorff & Bornmann, 2011) which combines normalized impact and output and the citing-side normalization where each citation is weighted with the citation density of the field in which the citing paper was published (Zitt & Small, 2008). The weighted citations are summed to calculate the normalized score for a research unit (e.g. an institution). It would be interesting to study the validity of summed NCs in comparison with other bibliometric approaches by correlating them, for example, with the assessment of papers by peers (Bornmann & Marx, 2015). (ii) It is standard in bibliometrics to normalize citation counts. Abramo and D'Angelo (in press) argue that not only citation impact but also output should be normalized. Many studies have shown that there are different intensities of publication across the fields (see e.g. Marx & Bornmann, 2015) and scientometricians could develop methods to normalize output scores. Following the impact normalization in bibliometrics, the methods should result in scores on the level of single publications (e.g. by dividing every publication by the publication density in a field).